\documentclass{pasj00}

\begin{document}
\SetRunningHead{T Enoto et al.}{Suzaku observation of an activated magnetar 1E 1547.0$-$5408}

\title{Suzaku Discovery of a Hard X-Ray Tail \\
 in the Persistent Spectra from the Magnetar \\
 1E 1547.0$-$5408 during its 2009 Activity}



%

\author{
Teruaki \textsc{Enoto},\altaffilmark{1}
Kazuhiro \textsc{Nakazawa},\altaffilmark{1}
Kazuo \textsc{Makishima},\altaffilmark{1,2}  \\
Yujin E. \textsc{Nakagawa},\altaffilmark{2}
Takanori \textsc{Sakamoto},\altaffilmark{3}  
Masanori \textsc{Ohno},\altaffilmark{4} 
Tadayuki \textsc{Takahashi},\altaffilmark{4} \\
Yukikatsu \textsc{Terada},\altaffilmark{5} 
Kazutaka \textsc{Yamaoka},\altaffilmark{6} 
Toshio \textsc{Murakami},\altaffilmark{7}
and 
Hiromitsu \textsc{Takahashi}\altaffilmark{8}
}
\altaffiltext{1}{
Department of Physics, University of Tokyo, 
7-3-1 Hongo, Bunkyo-ku, Tokyo, 113-0033}
 \email{enoto@juno.phys.s.u-tokyo.ac.jp}
 \altaffiltext{2}{
 Cosmic Radiation Laboratory, 
Institute of Physical and Chemical Research, \\
Wako, Saitama, 351-0198}
\altaffiltext{3}{
GoddardSpaceFlight Center, NASA, 
Greenbelt, Maryland, 20771, USA}
\altaffiltext{4}{
Department of High Energy Astrophysics, Institute of Space and Astronautical Science, \\
Japan Aerospace Exploration Agency, 3-1-1 Yoshinodai, Sagamihara, Kanagawa 229-8510}
\altaffiltext{5}{
Department of Physics, Saitama University, 
255 Shimo-Okubo, Sakura, Saitama 338-8570
}
\altaffiltext{6}{
Department of Physics and Mathematics, Aoyama Gakuin University, \\
5-10-1 Fuchinobe, Sagamihara, Kanagawa 229-8558, Japan}
\altaffiltext{7}{
Department of Physics, Kanazawa University, 
Kakuma, Kanazawa, Ishikawa 920-1192}
\altaffiltext{8}{
Department of Physical Science, Hiroshima University, \\
1-3-1 Kagamiyama, Higashi-Hiroshima, Hiroshima 739-8526
}


\maketitle

\begin{abstract}
The fastest-rotating magnetar 1E 1547.0$-$5408 was observed 
	in  broad-band X-rays with  Suzaku
	for 33 ks  on 2009 January 28--29,
7 days after the onset of its latest bursting activity.
After removing burst events, 
	the absorption-uncorrected 
	2--10 keV flux of the persistent emission was measured with the XIS
	as $5.7\times 10^{-11}$ ergs cm$^{-2}$ s$^{-1}$, 
	which is 1--2 orders of magnitude higher than was measured 
	in 2006 and 2007 when the source was less active.
The persistent emission was also detected significantly with the HXD 
	in $>10$ keV up to  at least $\sim$110 keV,
	with an even higher flux of $1.3\times 10^{-10}$ ergs cm$^{-2}$ s$^{-1}$ in 20--100 keV.
The pulsation was detected at least up to 70 keV
	at a  period of $2.072135 \pm 0.00005$ s, 
	with a deeper modulation than was measured in a fainter state.
The phase-averaged 0.7--114 keV spectrum 
	was reproduced by an absorbed blackbody emission 
	with a temperature of $0.65\pm0.02$ keV,
	plus a hard power-law with a photon index of  $\sim1.5$.
At a distance of 9 kpc,
	the bolometric luminosity of the blackbody
	and the 2--100 keV luminosity of the hard power-law are estimated as 
	$(6.2 \pm 1.2) \times 10^{35}$ ergs s$^{-1}$ and 
	$1.9\times 10^{36}$ ergs s$^{-1}$,  respectively,
	while the blackbody radius becomes $\sim 5$ km.
	Although the source had not been detected significantly
	in hard X-rays during the past fainter states,
	a comparison of the present and past spectra in energies below 10 keV suggests
	that the hard component is more enhanced than the soft X-ray component
	during the persistent activity.
\end{abstract}

\section{Introduction}
Luminous neutron stars can be classified into three categories
	from their dominant energy sources;
	i.e., rotation-powered, accretion-powered, and magnetic-powered objects.
The 1st and 2nd groups are represented by radio pulsars and 
	mass-accreting neutrons-star binaries, respectively.
The last category comprises those objects which are called magnetars
	\citep{duncan92,thompson95,thompson96}, 
	mainly observed in X-rays.
Magnetars have rotation periods of $P=2-12$ s 
	and  period derivatives of $\dot{P} \sim 10^{-11}$ s s$^{-1}$ 
	 \citep{woods06,mereghetti08}.
Their slow rotations cannot afford 
	their high soft X-ray luminosities, $\sim 10^{35}$ ergs s$^{-1}$;
	nor are they accretion-powered objects
	since they lack evidence of companion stars 
	(e.g., \cite{koyama89}).
Further considering their bursting activity and 
	strong surface magnetic fields ($>4.4\times 10^{13}$ G)
	evaluated from their $P$ and $\dot{P}$ values,
	the energy source of magnetars is considered to be their 
	strong magnetic fields.
Their	 X-ray luminosities, 
	which generally exceed their spin-down energy losses,
	can be explained by assuming that 
	their magnetic energies are released faster than their rotational energies.

Currently,
	$\sim$5 Soft Gamma-ray Repeaters (SGR)
	and $\sim$10 Anomalous X-ray Pulsars (AXP) 
	are known as magnetars\footnote{
	http://www.physics.mcgill.ca/~pulsar/magnetar/main.html
	}.
The former objects sometimes emit giant flares
	with peak luminosities up to $\sim$$10^{46}$ ergs s$^{-1}$,
	while the latter were discovered  by their bright soft X-ray ($\lesssim$10 keV)
	emission with luminosities of (0.1--1)$\times 10^{35}$ ergs s$^{-1}$.
Recent studies suggest that SGRs and AXPs are essentially the 
	same class of objects, 
	and their distinction is not fundamental \citep{mereghetti09b};
	i.e., 
	these names reflects  mainly their ways of discovery.

One distinguishing property of magnetars,
	observed not only from SGRs \citep{israel08,enoto09}
	but also from AXPs \citep{gavriil02,kaspi03},
	is occasional periods of high activity wherein numerous ``bursts" are emitted.
These bursts individually have  a typical duration of a few 
	hundred milliseconds and an energy release of $\gtrsim$$10^{37}$ ergs.
They are thought to represent sudden release of the magnetic energy,
	leading to the production of 
	hot plasmas and/or energetic particles somewhere in the system.

Bright persistent soft X-rays have long been observed 
	from magnetars in energies below $\sim$10 keV.
Empirically, these spectra have been fitted by a two-blackbody model 
	or a blackbody (of temperature $kT$$\sim$0.3--0.6 keV) 
	plus a soft power-law (of photon index $\Gamma=2$--$4$) model,
	where the power-law is thought to  
	arise via some sort of Compton process \citep{TLK02,rea09}.

Recently, INTEGRAL discovered from 
	several magnetars a new emission component 
	\citep{kuiper04,kuiper06,gotz06}.
	Becoming prominent in energies above $\sim 10$ keV,
	this component exhibits a high pulsed fraction (sometimes $\sim100\%$ at $\sim100$ keV) 
	and extends to $\sim100$ keV or more with a  surprisingly 
	hard photon index of $\Gamma \sim$1.
Although this discovery aroused wide interest, 
	the origin of this hard X-ray component is still an open question
	(e.g., \cite{heyl05, beloborodov07,baring07}).

So far, the hard tail has been detected from
	three SGRs (SGR~1900+14, SGR~1806-20, and SGR~0501+4516;
	e.g., \cite{gotz06, esposito07,rea09}) and 
	three AXPs (4U~0142+61, 1RXS~J170849.0-400910, and 1E~1841-045;
	e.g., \cite{kuiper04,kuiper06}).
While some detections from SGRs were made in their burst-active states,
	those from AXPs are limited to their quiescence.
It is hence unknown how the hard tails
	of AXPs behave in their active periods.
When attempting to answer these questions,
	Suzaku \citep{mitsuda07} provides a great advantage,
	because it allows us to simultaneously detect the  
	two spectral components
	of these objects in rather short exposures.

The X-ray source 1E~1547.0$-$5408 was discovered by 
	the Einstein Observatory in 1980 \citep{lamb81},
	and was suggested by recent X-ray observations as a magnetar candidate
	associated with a young supernova remnant G327.24-0.13 \citep{gelfand07}.
Subsequent radio observations 
	discovered pulsations with $P=2.069$ s 
	and  $\dot{P}=(2.318\pm0.005)\times 10^{-11}$.
This established 1E~1547.0$-$5408 as the fastest rotating magnetar
	known to date,
	with a surface field strength of $2.2\times 10^{14}$ G,
	a characteristic age of 1.4 kyr, 
	and a spin-down luminosity of $1.0\times 10^{35}$ ergs s$^{-1}$,
	all estimated from the measured $P$ and $\dot{P}$
	\citep{camilo07}. 
The radio observation also gave a distance estimate as  9 kpc 
	from its dispersion measure,
	and identified 1E~1547.0$-$5408 (PSR J1550-5418) 
	as a second example of transient radio magnetars
	after XTE J1810-197 \citep{camilo06}.
Since then, X-ray monitoring observations detected 
	a possible X-ray outburst in 2007 \citep{halpern08}.
The X-ray pulsation, which was not confirmed in quiescence \citep{gelfand07},
was first detected with XMM-Newton during this enhanced activity \citep{halpern08}.

Following a possible precursor phase in 2008 October
	\citep{Krimm2008GCN8311,Krimm2008GCN8312},
	the Swift Burst Alert Telescope detected bursting activity from 1E~1547.0$-$5408
	on 2009 January 22
	(Gronwall et al. 2009). 
A large numbers of short bursts were recorded by several X-ray satellites,
	including INTEGRAL \citep{savchenko09,mereghetti09a}, 
	Fermi \citep{connaughton09}, 
	Konus-Wind \citep{golenetskii09}, 
	and RHESSI \citep{bellm09}.
The Wide-band All-sky Monitor (WAM) on board Suzaku 
	also recorded $\sim$250 short bursts on January 22 \citep{terada09}.
Based on this information, 
	a Suzaku Target-of-Opportunity (ToO) observation was conducted on January 28.
In the present paper, 
	we report on the result of this observation,
	focussing on wide-band spectra of the persistent emission.
Analysis of short bursts will be reported elsewhere.

\section{Observation}
\label{Observation}
The present ToO observation of 1E 1547.0$-$5408 was carried out with Suzaku
	from 2009 January 28 21:34 (UT), until January 29 21:32 (UT),
	with a gross duration of $\sim$86 ks.
This was 7 days 
	after the onset of its latest strong bursting activity 
	on January 22 \citep{gronwall09}.
The Suzaku observation partially overlapped with those by 
	INTEGRAL, RXTE, Swift, and Chandra.

Of the three operating cameras of the X-ray Imaging Spectrometer (XIS; \cite{koyama07}),
XIS1 (a back-illuminated or BI CCD) and XIS3 (a front-illuminated of FI CCD) were operated 
	incorporating 1/4 window option to ensure a time resolution of 2 s.
Burst option was also employed to avoid possible photon pile-up problems,
	with its exposure reduced to 1/4 (i.e., 0.5 s exposure and 1.5 s artificial dead-time),
	although the actual persistent flux of 1E 15470-5408 was 
	low enough to allow an operation without this option.
The other one, XIS0, was operated in the timing mode (P-sum mode),
	which provides a $\sim 7.8$ ms time resolution, 
	together with only 1-dimentionally projected position information.

The Hard X-ray Detector (HXD; \cite{takahashi07}) 
	was operated in the standard mode,
	wherein individual events are recorded with a time resolution of 61 $\mu$s.
The target was placed at the ``HXD nominal" position.

\section{Data Reduction}
\label{Data Reduction}
Using version 6.6.2 HEADAS software,
	we analyzed the Suzaku data of 1E~1547.0-5408 (OBSID = 903006010)
	prepared through version 2.3.12.5 processing.
Events were discarded if they were acquired in the South Atlantic Anomaly (SAA) or 
	in regions of low cutoff rigidity ($\le$6 GV for the XIS and $\le$8 GV for the HXD),
	or with low Earth elevation angles (ELV$<$5).
The net exposures were 42.6 ks, 10.6 ks, 10.7 ks, and 33.5 ks 
	for XIS0, XIS1, XIS3, and the HXD, respectively.
The reduced exposures of XIS1 and XIS3 are 
	due to the burst option (section \ref{Observation}).

\subsection{XIS data reduction}
\label{XIS-reduct}

For our spectroscopic studies,
	we use the data from XIS1 and XIS3, 
	which were acquired in the normal imaging mode.
The source was detected clearly in the XIS images.
On-source XIS1 and XIS3  events were extracted 
	from a region of $2'$ in radius, centered on the source centroid.
Background events were derived from a similar region 
	as far away from the source as possible.
After  subtracting this  background,
	we detected 1E~1547.0$-$5408 
	at a burst-inclusive  2--10 keV average count rate
	as given in table~\ref{tbl:counts}.
The higher XIS3 rate is due to a relatively hard spectrum and 
	a strong absorption (subsection \ref{Persistent phase-averaged spectrum}),
	coupled with a more hard-band enhanced efficiency of  FI CCDs
	than that of a BI chip.

The data from XIS0, taken in the timing mode,
	are used in our timing analyses.
We applied some additional corrections to the XIS0 data.
Namely, we further filtered the cleaned event file 
	using a selection criteria as 
	\verb+((GRADE==0)||(GRADE==1)||(GRADE==2))+, 
	and eliminated hot pixels and flickering pixels.
In addition,
	we shifted nominal time assignments 
	of the XIS0 events by $-60$ ms (i.e., \verb+TIME-0.06+ s),
	to take into account the following two effects.
One is the read-out time lag
	when a target is placed at the HXD nominal position \citep{koyama07}.
The other is a $30\pm16$ ms offset
	of the absolute timing of the XIS data acquired in the P-sum mode
	(Matsuda et al., 2009).
This offset was measured by observing 
	the Crab pulsar 
	and 
	Her X-1,
	and 
	comparing the results with simultaneous HXD data 
	of which the relative and absolute timing accuracies 
	had already been verified \citep{terada08}.
These observations of the two pulsars also confirmed the relative timing accuracy of the XIS.	

\subsection{HXD-PIN data reduction}
\label{PIN-reduct}
\begin{figure}
\begin{center}
\FigureFile(80mm,){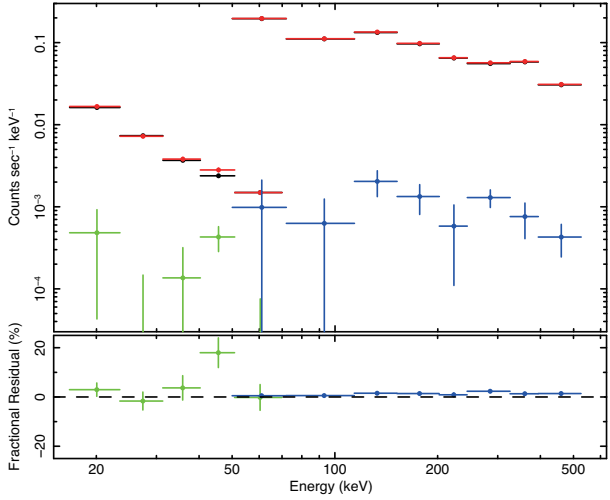}
\caption{
(Top) A comparison between the Earth occulted HXD-PIN and HXD-GSO data (red) 
and
the corresponding simulated background data (black).
The residuals (data minus simulation) are shown in green
for HXD-PIN and in blue for HXD-GSO.
(Bottom) Fractional residuals, namely (data - simulation) / simulation.
}
\label{fig:figure1.eps}
\end{center}
\end{figure}

From the HXD-PIN data,
	we subtracted estimated non X-ray background (NXB),
	which were produced by the ``tuned" (LCFITDT) NXB model \citep{fuka09}.
In order to assess the accuracy of the background modeling,
	we compared its prediction with Earth-occulted portion of the data with 14 ks exposure,
	because these data are considered to consist solely of the NXB.
These two spectra are compared in figure \ref{fig:figure1.eps}.
On average, the simulated background was found to fall below the actual Earth 
	occultation data by 2.3\% in the 15--70 keV range.
The discrepancy is relatively large (13.0\%) in the 40--50 keV range, 
	presumably because this band is affected
	 by instrumental Gd-K lines (43.0 keV, 42.3 keV, 48.6 keV, and 50 keV).
However, 	 
	if the 40--50 keV range is ignored, 
	the data vs. model difference reduces to only 1.0\%.
Therefore, we consider the HXD-PIN NXB modeling to be appropriate,
	and assign  it a systematic error of 2.0\% after \citet{fuka09}.


From the HXD-PIN data, we also subtracted
	an expected contribution from the Cosmic X-ray Background (CXB).
	This was calculated using a recent result from \citet{moretti09},
	which describes the CXB photon number spectrum as
\[
\frac{C \cdot \Omega }{(E/E_0)^{\Gamma_1}+(E/E_0)^{\Gamma_2}} \\
\  \textrm{photons s$^{-1}$ cm$^{-2}$ keV$^{-1}$~. }
\]
Here,
	$C=1.33\times 10^{-4}$ is a normalization,
	$E_0=29$ keV is a characteristic energy, 
	$\Gamma_1=1.40$ and $\Gamma_2=2.88$ are photon indices,
	and $\Omega=1.2\times 10^{-3}$ str is the solid angle
	of an assumed 2\degree $\times$ 2\degree emission region
	with uniform brightness, 
	corresponding to the HXD-PIN ``flat" response.
The CXB counts are subject to 
	$\sim11\%$ sky-to-sky fluctuations \citep{fuka09}.
The average CXB flux corresponds to $\sim5\%$ of PIN-NXB.

Since 1E 1547.0$-$5408 is located on the Galactic plane 
	at Galactic coordinates of $(l,b)=(-32\degree .8, -0 \degree .1)$,
	Galactic Ridge X-ray Emission (GRXE) may contaminate as well.
INTEGRAL mapping observations showed 
	that the typical 17--60 keV GRXE flux is less than 
	$1\times10^{-9}$ erg s$^{-1}$ cm$^{-2}$ withn an IBIS FOV,
	on the Galactic disk at $|l|>20$\degree (figure 13 of \cite{krivonos07}).
Within the HXD-PIN FOV,
	we then expect a flux of $\lesssim 2\times10^{-11}$ erg s$^{-1}$ cm$^{-2}$,
	which is less than $\sim$1\% of NXB or $\sim$20\% of CXB.
This estimate is also supported by a near-by blank sky observation (OBSID = 501043010)
	at  $(l,b)=(-29\degree .6, -0\degree. 38)$, 3\degree.2 off  1E 1547.0$-$5408:
	after subtracting the nominal NXB and CXB, 
	the GRXE contribution in the 12--70 keV band  was $\sim$30\% of the average CXB.
	Therefore,  the GRXE contribution in the HXD-PIN data is negligible.

As summarized in table~\ref{tbl:counts},
	we were left with positive signals of $\sim 0.174\pm 0.004$ cts s$^{-1}$ 
	in the 15--70 keV HXD-PIN band,
	when the NXB and CXB were both subtracted.
	Thus, the signal detection is highly significant in the HXD-PIN data,
	because the systematic error (2.0\%) 
	is estimated to be $\sim 7\times 10^{-3}$ cts s$^{-1}$ in the same band.

\subsection{HXD-GSO data reduction}
\label{GSO-reduct}
The simulated background of HXD-GSO, also produced by the ``tuned" model,
	was evaluated again by the Earth-occulted data.
	As shown in figure \ref{fig:figure1.eps},
	the simulated background systematically under-predicted
	the actual Earth occulted HXD-GSO data by 
	0.53\% and 1.4\% in the 50--114 and 115--578  keV band, respectively.
Conservatively,
	we hence subtracted the HXD-GSO background 
	after increasing it by 0.53\% and 1.4\% over the above ranges, respectively,
	and assigned systematic uncertainties of
	0.60\% and 0.59\% therein \citep{fuka09}.

After subtracting the NXB and considering the above systematic errors,
	the HXD-GSO signal count rate became as shown in table~\ref{tbl:counts}.
Therefore, 
	the signal has an overall significance of 
	$\sim$2.3$\sigma$ in the 50--114 keV range,
	but is insignificant in  higher energies.
\subsection{Elimination of burst events}
\label{burst-elim}

\begin{figure*}[thb]
\begin{center}
    \FigureFile(130mm,){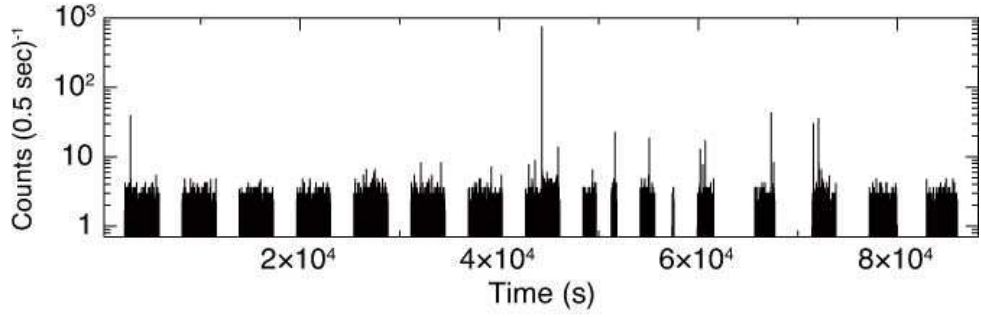}
\end{center}
\caption{
A 2.0-s (with the artificial 1.5-s dead times) binned 2--10 keV light curve of the entire observation, 
obtained by XIS1 and XIS3, shown after  background subtraction.
}\label{fig:lc}
\end{figure*}

Figure \ref{fig:lc} shows a dead-time corrected 
	and background-subtracted XIS1+XIS3 light curve
	in the 2--10 keV range.
	There, a number of short burst events can be noticed as sharp spikes.
By visually inspecting this XIS1+XIS3 light curve
	and  that of HXD-PIN (15--70 keV),
	produced with a bin width of 2 s (one frame) and 1 s, respectively,
	we identified 13 prominent burst candidates.
Some of them were detected by both instruments, 
	while the others by  either of them. 
	Then, the XIS and HXD data acquired within  3-s time intervals 
	both before and after each of these bursts were discarded.

After this burst elimination, 
	the background-subtracted average count rates 
	of  XIS, XIS3, and HXD-PIN became as given in table ~\ref{tbl:counts}.
Although the remaining data must still contain smaller bursts,
	their summed contribution is estimated to be no more than 0.1\%.

\subsection{Contaminating sources}
During the Suzaku observation,
	the 10--30 keV HXD-PIN signal intensity,
	after subtracting the NXB and CXB, was $\sim$10 mCrab.
According to the INTEGRAL General Reference Catalog ({\bf gnrl\_refr\_cat\_0030.fits}),
	the HXD-PIN full FOV ($\sim 34' \times \sim 34'$)
	contained 
	4 contaminating source candidates 	
	detected with ASCA \citep{sugizaki2001}.
Their 10--30 keV intensities (using JEM\_X),
	0.01 mCrab of AX J1550.5-5408,
	0.26 mCrab of AX J1549.8-5416 ,
	0.01 mCrab of AX J1553.5-5347,
	and 
	0.41 mCrab of AX J1549.0-5420,
	sum up to $<$10\% of the signal from the 1E~1547.0$-$5408 field.
Although these may be variable X-ray sources,
	AX~J1550.5-5408 was not detected neither with 
	the XIS, nor in a 3--10 keV Swift/XRT image
	obtained on 2009 January 29.
The Swift/XRT count rate of AX J1549.8-5416 was $\lesssim$0.01 times 
	that of 1E~1547.0-5408 on the same occasion.
In addition,
	AX J1553.5-5347 and AX J1549.0-5420 were located 
	outside the FWHM FOV of HXD-PIN,
	with their aperture transmissions of 
	$\lesssim$10\% and $\lesssim$50\%, respectively.
Further considering lack of reports on X-ray brightening from none of these sources,	
	we regard their contamination as negligible.
This assumption is further confirmed in section 4.2.	
	

As summarized in table \ref{tab:cont_list},
	the $4^\circ \times 4^\circ$ HXD-GSO FOV (and outside that of HXD-PIN)
	contains 5 sources with 20--60 keV intensities of  $>$3 mCrab.
	Since they are also variable X-ray sources,
	we examined the RXTE/ASM quick-look data \citep{jahoda96}  
	for their intensities over 10 days around our Suzaku observation.
All of them were found, during this period,
	to be $<$20 mCrab in the 2--10 keV range.
When we assume spectral shapes 
	from the INTEGRAL General Reference Catalog
	and normalizations of the ASM monitoring,
	their 50--200 keV intensities are estimated to be  less than $\sim$0.01\% level of the HXD-GSO NXB.
Therefore we can also ignore source confusion within the HXD-GSO FOV.
From these examinations, 
	we regard the burst-removed signals 
	detected with HXD-GSO (below 114 keV),
	given in the last column of table~\ref{tbl:counts},
	as coming also from 1E~1547.0$-$5408 itself.

\section{Data Analysis and Result}

\subsection{Pulsation}
\label{pulsations}
\begin{figure}[hbt]
\begin{center}
      \FigureFile(90mm,){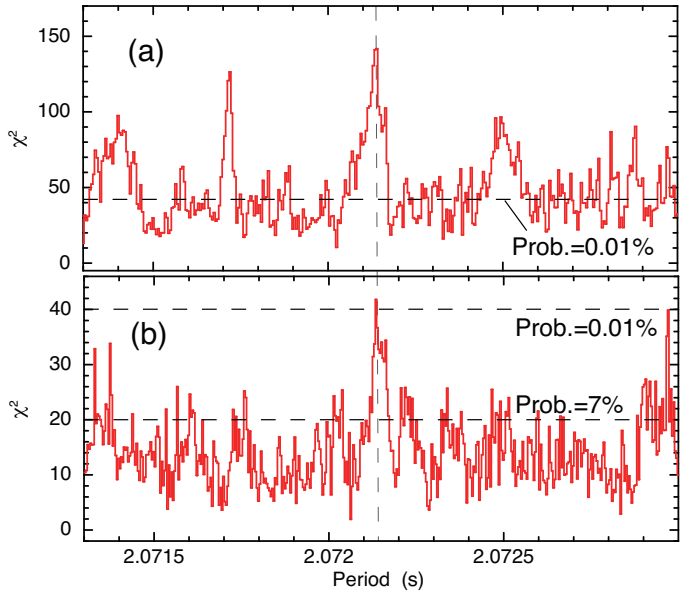}
\caption{
Periodograms derived with an epoch-folding analysis from
	the 2--10 keV XIS0 P-sum mode data (panel a)
	and 
	the background-inclusive 12--70 keV HXD-PIN data (panel b).
The number of phase bins per period 
	is indicated by 13 and 15 for the XIS and HXD-PIN, respectively.
Typical chance probabilities are horizontal lines.
Sub-peaks at 2.0717 s and 2.0725 s in panel a 
	correspond to beats with half the spacecraft orbital period ($\sim$45 min).
}
\label{fig:periodgrams}
\end{center}
\end{figure}

After eliminating the 13 burst candidates,
	all the XIS0 and HXD event arrival times were converted to 
	those measured at the solar system barycenter,
	using a source position of 
	$(\alpha, \delta)_{2000}=(15^{\rm h}50^{\rm m}54.11^{\rm s}, $-$54\degree18'23.7'')$
	as well as the spacecraft orbital information.
Then, via standard epoch-folding technique,
	we  searched the 2--10 keV XIS P-sum data 
	and the 12--70 keV background-inclusive HXD-PIN data 
	for the known 2.0 s X-ray period \citep{halpern08}.
The analysis was performed over a trial period range of 2.0713--2.0730 s,
	with a step of $4\times 10^{-6}$ s 
	which corresponds to a change of 0.08 cycle across the overall time span 
	of $\sim$86 ks.
The derived periodograms are shown in figure~\ref{fig:periodgrams}.
A significant 
	($\chi^2/\nu=41.8/12$ for the HXD-PIN data, and 
	$\chi^2/\nu=141.1/14$ for the XIS P-sum data)
	periodicity was found at a consistent period of 
\begin{equation}
       P=2.072135\pm0.00005 \ \mathrm{s}
\label{eq:period}
\end{equation}
	as of epoch 54859 (MJD).
The analysis employed a period derivate $\dot{P}=2.35\times 10^{-11}$ s s$^{-1}$,
	from multi-satellite timing observations of 1E~1547.0$-$5408 (Israel et al. private communication)
	which covers our Suzaku observation epoch.
We use equation  (\ref{eq:period}) and $\dot{P}=2.35\times 10^{-11}$ s s$^{-1}$ 
	for our subsequent timing analyses.

\begin{figure}[hbt]
\begin{center}
      \FigureFile(70mm,){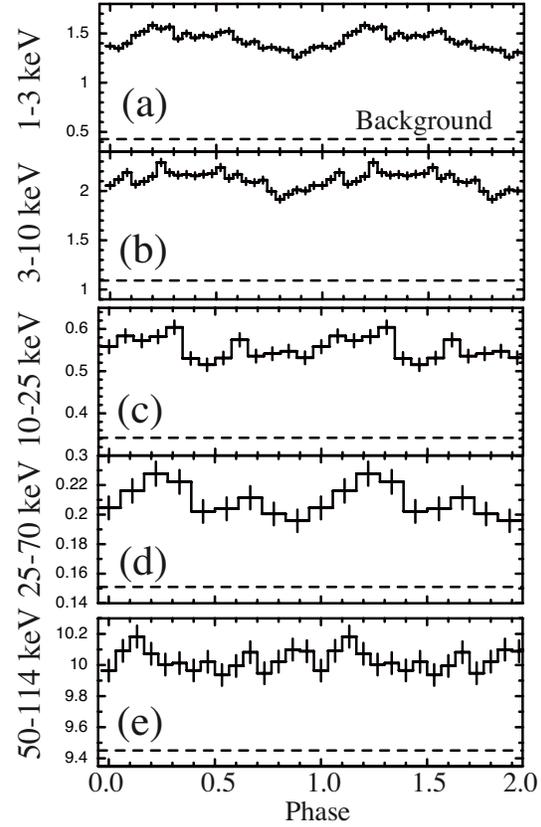}
\caption{
Energy-sorted pulse profiles, 
	folded at the period of equation 1. 
	Two pulse cycles are displayed for clarify,
	and the background level is indicated as dashed lines.
Panels (a) and (b) are extracted from XIS0 (timing mode) 
	in the 1--3 and 3--10 keV energy ranges, respectively,
	while panels (c) and (d) are from HXD-PIN in the 15--30 
	and 30--70 keV energy ranges, respectively.
Panel (e) is from the 50--114 keV GSO data.
The significance of the pulsation is 
	given in the text.
\label{fig:profiles}
}
\end{center}
\end{figure}

Figure \ref{fig:profiles} shows
	energy-sorted pulse profiles  derived from XIS0, HXD-PIN, and HXD-GSO,
	folded at equation  (\ref{eq:period}).
The null hypothesis probability of the pulse detection is 
	$< 1\times10^{-4}$, $< 1\times10^{-4}$, $3\times 10^{-4}$,  0.06, and 0.47,
	in the 1--3 keV (XIS0), 3--10 keV (XIS0), 10--25 (HXD-PIN), 
	25--70 (HXD-PIN), and 50-114 keV (HXD-GSO)  energy bands, respectively.
Therefore, the pulsations are significant in the XIS and the low-energy ($\lesssim$25 keV) 
	HXD-PIN data with $>$3 sigma,
	although 
	inconclusive in the high-energy HXD-PIN band 
	and 
	insignificant in the HXD-GSO data.
The HXD-PIN data were further analyzed with	
	the Z-squared tests \citep{1983A&A...128..245B,1994MNRAS.268..709B}.
Then, 
	the 10--25 keV and 25--70 keV HXD-PIN data yielded 
	$Z_2^2=19.7$ (the total event number being $N\simeq 1.7\times 10^4$) and $12.7$ ($N\simeq 6.5\times 10^4$),
	which correspond to null hypothesis probabilities of $6\times 10^{-4}$ and $0.013$, respectively.
In other words, 
	the high-energy (25--70 keV) HXD-PIN data are modulated,
	with a $\sim$99\% confidence,
	at the same period as the lower-energy photons,
	even though the $\chi^2$ significance is not high enough.
Therefore,
	we discuss the 25--70 keV pulse profile as well.

\begin{figure}[hbt]
\begin{center}
\FigureFile(70mm,40mm){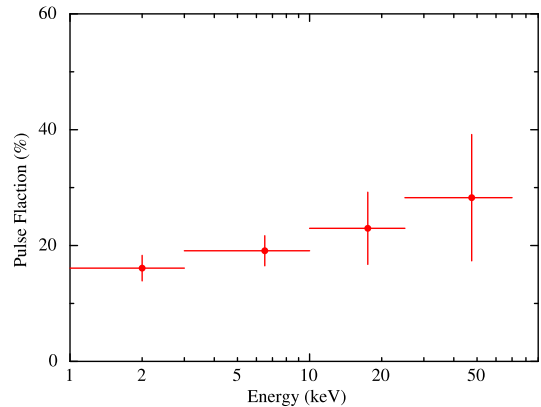}
\caption{
The pulsed fraction shown as a function of energy.
}
\label{fig:fractions}
\end{center}
\end{figure}

As shown by figure~\ref{fig:profiles}  and reported before \citep{halpern08},
	the pulse profiles of this object are relatively shallow.
The peak-to-peak pulsed fractions, defined as 
	$(F_{\mathrm{max}} - F_{\mathrm{min}})/(F_{\mathrm{max}} + F_{\mathrm{min}})$,
	are presented in figure~\ref{fig:fractions}
	as a function of the photon energy.
Here, $F_{\mathrm{max}}$ and $F_{\mathrm{min}}$ 
	are the maximum and minimum background-subtracted
	count rates across the pulse phase, 
	respectively.
Error bars include
	Poisson fluctuations and 
	the systematic uncertainties of the NXB.
Although the pulsed fraction tends to increase towards higher energies
	as observed in some magnetars \citep{kuiper06},
	the dependence is insignificant when considering the errors.

\subsection{Persistent phase-averaged spectrum}
\label{Persistent phase-averaged spectrum}

\begin{figure*}[htb]
\begin{center}
\FigureFile(100mm,80mm){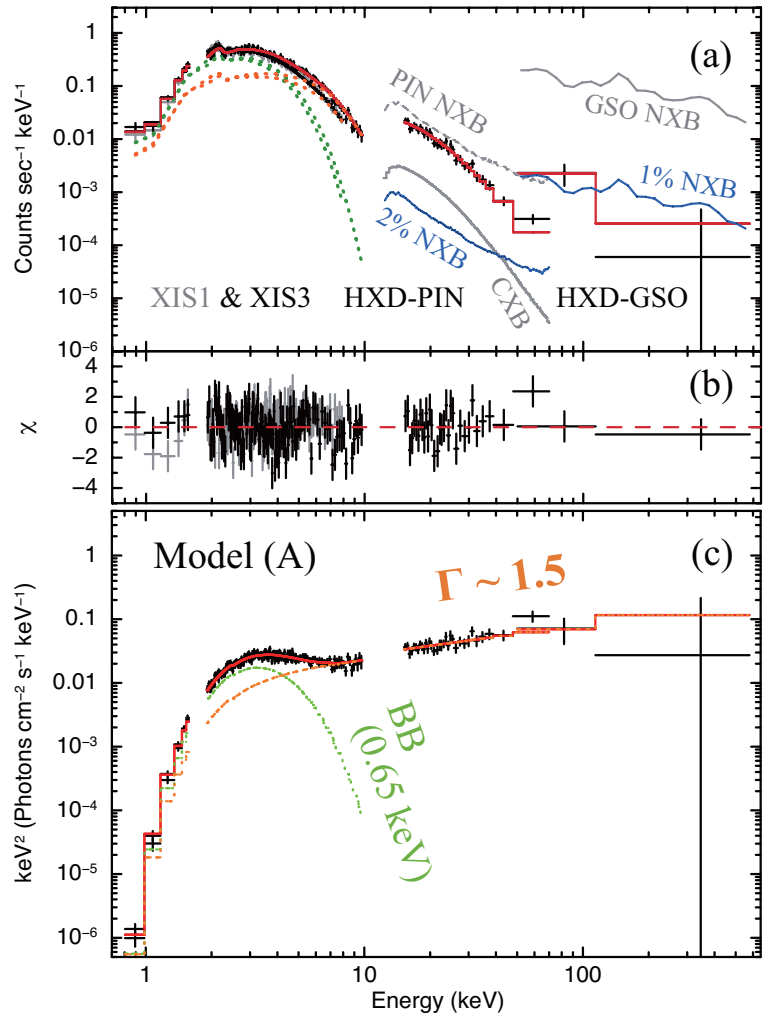}
\caption{
(a) Persistent phase-averaged X-ray spectra of 1E~1547.0$-$5408,
fitted simultaneously with a model consisting of a blackbody (green)
and a hard power-law (red).
Related background spectra are also shown.
(b)
Fit residuals.
(c)
An $\nu F_{\nu}$ form of panel (a) using Model (A).
}
\label{fig:bb_pl_result}
\end{center}
\end{figure*}

Figure \ref{fig:bb_pl_result}a shows time-averaged 
	and dead-time-corrected  spectra of the persistent emission from 1E 1547.0$-$5408,
	derived over a broad energy band with XIS1, XIS3, HXD-PIN, and HXD-GSO.
The data accumulation and background subtraction
	were carried out as described in sections~\ref{XIS-reduct}, sections~\ref{PIN-reduct},
	and sections~\ref{GSO-reduct}.
The 13 burst events were excluded (section~\ref{burst-elim}).
In addition, as described in sections~\ref{GSO-reduct},
	the 0.6\% systematic errors were assigned to the HXD-GSO data, 
	while  the 2.0\% systematic error in the HXD-PIN NXB
	(subsection~\ref{PIN-reduct}) is separately considered later.
In agreement with figure \ref{fig:profiles}e and the results in sections~\ref{GSO-reduct},
	the hard X-ray signals were detected with HXD-GSO
	at least up to 114 keV with 
	2.3 sigma level. 
This significance increases to $>$3$\sigma$ 
	when the upper bound energy is lowered to 70 keV of HXD-PIN.

Since the small pulsed fractions (figure~\ref{fig:fractions})
	and the short exposure (33 ks with the HXD) make it difficult 
	to perform detailed phase-resolved spectroscopy,
	hereafter we analyze only these phase-averaged spectra.
The analysis utilized  {\tt xspec} version 12.5,
	with the HXD response matrices version 2009-04-03,
	and the XIS response matrices created using {\bf xissimarfgen} and {\bf xisrmfgen}.

As clearly revealed by figure \ref{fig:bb_pl_result}c
	in an $\nu F_\nu$ spectral form,
	the detected HXD-PIN and HXD-GSO signals
	show a prominent hard component above 10 keV,
	just like in some other magnetars (e.g., \cite{kuiper06}).
This provides the first detection of such a hard spectral
	component from this magnetar.
In deriving figure \ref{fig:bb_pl_result}c,
	 the cross-normalization factor of HXD-PIN relative to the XIS 
	was fixed at 1.181 after Maeda et al (2008).
Since the XIS and the HXD-PIN spectra thus connect smoothly to each other,
	the HXD-PIN signals are unlikely to be contaminated by other sources 
	that are inside the HXD-PIN FOV but outside that of the XIS.
Any signal from such contaminants is estimated to be $\lesssim$24\% 
	at 15 keV of the HXD-PIN spectrum.
This level would be much lower in harder energies,
	considering that such sources, most likely low-mass neutron star binaries \citep{sugizaki2001},
	should exhibit softer spectra than the observed one.
	
In order to approximately characterize this hard component, 
	we fitted the 15--70 keV HXD-PIN data in figure \ref{fig:bb_pl_result}a 
	with a single power-law model.
The fit was successful with $\chi^2/\mathrm{d.o.f}=30.4/31=0.98$,
	and yielded a photon index of 
	$\Gamma=1.41^{+0.17}_{-0.16} (\mathrm{stat.})\pm 0.04 (\mathrm{sys.})$,
	and a 15--70 keV flux of 
	$F_{\mathrm{x}}=1.38_{-0.19}^{+0.21} (\mathrm{stat.})\pm0.09 (\mathrm{sys.})
	\times 10^{-10} \mathrm{ergs \ s^{-1} \ cm^{-2}}$.
Here, the  systematic errors were estimated by increasing 
	or decreasing the PIN-NXB by 2\% (subsection~\ref{PIN-reduct}).
The result implies that the hard component is rather featureless.


As a next step, we included the XIS1, XIS3, and HXD-GSO spectra.
The absorbing column density $N_{\rm H}$ was allowed to vary.
Then, as shown in figure  \ref{fig:bb_pl_result}
	and table \ref{tab:spec_phase_average}a,
	the entire 0.7--114 keV spectra were successfully fitted 
	by adding a blackbody with a temperature $kT=0.65$ keV
	to the flat power-law which was found with the HXD-PIN data.
Hereafter, we call this ``blackbody+powerlaw" modeling Model A.
While the errors in table 1 are statistical only,
	inclusion of the HXD-PIN systematic errors affected  $\Gamma$ by $\pm$0.02, 
	and the 20--100 keV flux by  $\pm$8\%.
The fit is reproduced in figure~\ref{fig:bb_pl_result}c in an $\nu F_\nu$ form.

\begin{figure*}[htb]
\begin{center}
\FigureFile(160mm,50mm){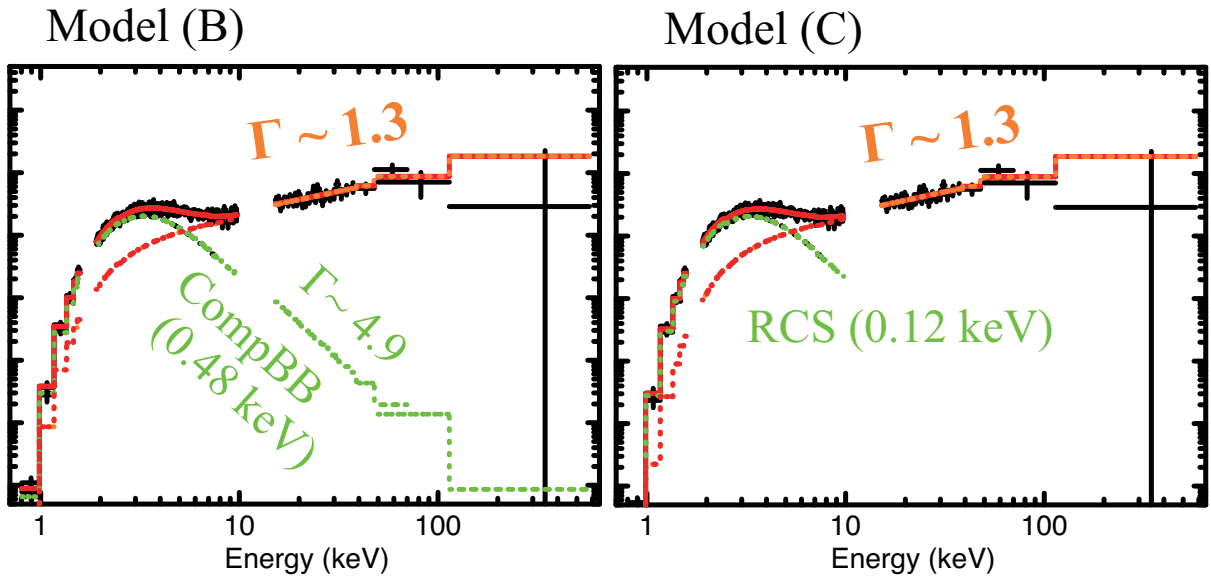}
\caption{
The same as figure \ref{fig:bb_pl_result}c,
	but using Model B (left) and 
	Model C (right).
}
\label{fig:persist_comp}
\end{center}
\end{figure*}

Although this Model A gave an acceptable result,
	we tried to replace the backbody component, accounting for the soft signals, 
	with a Comptonized blackbody model.
This is because a soft power-law tail with $\Gamma$=2--4,
	suggestive of Comptonization, often emerges 
	in the soft X-ray ($<10$ keV) spectra of magnetars \citep{mereghetti08}.
In order to take into account this effect,
	we constructed a simple formalism of the Comptonized 
	blackbody radiation model \citep{rybicki79,tiengo05,halpern08},
	which is  described in detail in Appendix.
In high energies this model approaches a power-law form
	with a steep photon index $\Gamma_{\mathrm{comp}}$,
	but it is free from an infrared divergence
	that plagues  a more conventional modeling of
	a blackbody plus a steep power-law.
As presented in figure~\ref{fig:persist_comp} (left) and table~\ref{tab:spec_phase_average},
	this Comptonized blackbody plus hard power-law model,
	hereafter called Model B, also gave an acceptable (and slightly better) fit to the data.
The  photon index of the soft tail was obtained as 
	$\Gamma_{\mathrm{comp}} \sim 4.9$ (table \ref{tab:spec_phase_average});
	that of the  hard power-law remained unchanged 
	from that of Model A within errors;
	and the blackbody temperature decreased to  $kT=0.48$ keV.

Alternatively, 
	to represent the above soft power-law component,
	we may also employ 1D semi-analytical 
	resonant Compton scattering model after \citep{rea09}
	using {\tt xspec12} local model ``RCS.mod".
This local model assumes repeated resonant cyclotron scatterings
	of the thermal radiation from the surface 
	by hot electrons in the neutron star magnetosphere.
The model, hereafter Model C, also gave an acceptable fit
	(figure~\ref{fig:persist_comp} right), 
	although it requires a large optical depth (table~\ref{tab:spec_phase_average}).

The hard-tail slope, $\Gamma \sim 1.5$, must steepen at some energies,
	in order for the hard-component luminosity not to diverge.
To obtain a constraint on such a high-energy steepening,
	we employed Model A, and multiplied its  hard power-law 
	with an exponential cutoff factor of the form $\exp \left(-E/E_{\rm cut} \right)$,
	where $E$ is the photon energy and $E_{\rm cut}$ is a parameter.
Then,
	the data gave a constraint as
	$E_{\rm cut} > 200$ keV
	at the 90\%-confidence limit.

\section{Discussion}
\subsection{Broad-band information}
\label{broadband}

The present Suzaku ToO observation of 1E 1547.0$-$5408,
	made 7 days after the onset of the 2009 January bursting activity,
	have allowed the first study of this fastest-rotating magnetar
	over an extremely broad band spanning two  orders of magnitude in energy.
The most important finding is the first discovery from this source
	of the very hard component,
	which was so far observed from some (though not all) other magnetars.
This component dominates the burst-removed persistent emission spectra 
	of 1E 1547.0$-$5408 in $\gtrsim 7$ keV,
	and extends at least up to $\sim$100 keV with $\Gamma \sim1.5$
	without evidence of prominent spectral features
	or steepening ($E_{\textrm{cut}}>200$ keV).

We  discovered that the hard X-rays are 
	pulsed at the same period as the soft X-rays,
	although the pulsed fractions
	are considerably smaller (figure~\ref{fig:fractions}) than those of typical magnetars.

As summarized in table~\ref{tab:spec_phase_average},
	the measured 20--100 keV flux ($1.3\times 10^{-10}$ erg s$^{-1}$ cm$^{-2}$)
	exceeds the unabsorbed 2--10 keV flux 
	($8.0\times 10^{-11}$ erg s$^{-1}$ cm$^{-2}$)
	by a factor of $R=1.57 \pm 0.15$.
(This factor becomes 2.1 if instead using the absoption-uncorrected
	2--10 keV flux in table~\ref{tab:spec_phase_average}.)
Therefore, at least during this active period,
	the emission appeared mainly in  hard X-ray energies.
If we separate the soft and hard components referring to Model A for simplicity,
	and employ a distance $d=9$ kpc as a fiducial value,
	the absorption-corrected bolometric luminosity 
	of the soft component (blackbody) becomes
$L_{\mathrm{BB}}=(6.2\pm1.2) \times 10^{35}  (d/9 \textrm{\ kpc})^2$ ergs s$^{-1}$, 
	whereas that of the hard component in the 2--100 keV band
	becomes $1.9\times 10^{36} (d/9 \textrm{\ kpc})^2$ ergs s$^{-1}$.

 \citet{enoto10} pointed out 
	that the 20--100 keV to 2--10 keV  flux ratio $R$ of magnetars,
	as introduced above,
	negatively correlates with their characteristic age $\tau_{\rm c}$.
During the present activity, 
	1E 1547.0$-$5408 exhibited  $R=1.57$  as derived above.
This ratio is higher than those of magnetars with older characteristic ages
	(typically AXPs;
	e.g., $R\sim$0.7 for 4U~0142+61 with $\tau_{\rm c}=70$ kyr),
	while lower than those of objects with younger characteristic ages
	(typically SGRs;
	e.g. $R \sim 2.8$ for SGR~1806$-$20 with $\tau_{\rm c}=0.2$ kyr).
Furthermore, the hard-tail slope of 1E~1547.0$-$5408 ($\Gamma \sim 1.5$)
	is in between those of typical AXPs ($\Gamma \sim 1.0$)
	and of typical SGRs ($\Gamma=1.8-3.1$; \cite{gotz06}).
This is reinforced when we plot the photon index of 
	1E~1547.0$-$5408 on figure 3 of \cite{Kaspi2010},
	which reports a correlation between the hard-band 
	photon index and the magnetic field strength.	
These properties, together with $\tau_{\rm c}=1.4$ kyr,
	make the activated 1E 1547.0$-$5408 
	an object in between the typical SGRs with young characteristic ages
	and the typical AXPs with older characteristic ages.

\subsection{Modeling of the soft component}
\label{softcomp}

Taking into account the hard tail component,
	the broad-band Suzaku spectra have been reproduced successfully
	by the three alternative modelings (Models A, B, and C) of the soft component.
In all cases, the hydrogen column density was obtained as 
	$N_{\mathrm{H}}=(3.2-4.0)\times 10^{22}$ cm$^{-2}$ (table \ref{tab:spec_phase_average}),
	which is consistent with the past XMM-Newton and Swift measurements
	in a much faint state in 2006--2007 \citep{halpern08}.
Although this $N_{\mathrm{H}}$ is 1.5--1.8 times higher
	than the Galactic HI column density in the direction of 1E~1547.0$-$5408, 
	$2.2\times 10^{22}$ cm$^{-2}$ \citep{dickey90},
	the difference can be explained away by including 
	the H$_2$ gas contribution as judged from CO observations \citep{halpern08}.

Although the three models all proved to be successful,
	Model B and Model C give better chi-squares than Model A
	(figure \ref{fig:persist_comp}, table \ref{tab:spec_phase_average}).
According to $F$-tests,
	chance probabilities of the fit improvement from Model A to Model B,
	and Model A to Model C, are 
	$1.2\times10^{-3}$ and $8.3\times10^{-3}$, respectively.
Therefore,  the hardest end of the soft component 
	is suggested to exhibit a power-law like extension 
	as found in the quiescent spectrum of this object \citep{halpern08},
	rather than turning over exponentially.

\subsection{Effects of the enhanced activity}
\label{activity}

While 1E~1547.0$-$5408 showed  AXP-like soft X-ray spectra 
	in quiescence \citep{gelfand07,halpern08},
	its burst properties recorded in the present activity are 
	very similar to those of typical SGRs (Mereghetti et al. 2009).
Therefore, this object is considered to exhibit 
	properties typical of both SGRs and AXPs,
	depending on its activity states.
	It is hence expected to provide us a unique opportunity 
	in comparing magnetars of different activity levels.

Past Swift and XMM-Newton observations of this object reported 
	a historical minimum 1--8 keV flux (absorption uncorrected) of 
	$3.3\times 10^{-13}$ ergs cm$^{-2}$ s$^{-1}$  in  2006 August, 
	while a much enhanced value of $4.6\times 10^{-12}$ ergs cm$^{-2}$ s$^{-1}$
	during the previous outburst in 2007 January  \citep{halpern08}.
The  same quantity measured in the present observation, 
	$(5.2\pm0.2)\times 10^{-11}$ ergs cm$^{-2}$ s$^{-1}$ (using Model A for simplicity), 
	is  even $\sim$158 and $\sim$11 times higher than 
	those measured in 2006 August and  2007 January, respectively.
These large soft X-ray variations are similar to those of  the
	transient magnetars XTE J1810$-$197 and  CXOU J164710.2$-$455216 \citep{halpern08}.

In evaluating how the activity affected the soft component,
	let us for simplicity refer to Model A,
	even though the soft component may not be
	a pure blackbody  (section~\ref{softcomp}).
In terms of Model A,
	we measured a blackbody temperature of $0.65\pm0.02$ keV
	(table \ref{tab:spec_phase_average}),
	which is $\sim 1.5$ times higher  than the values of $0.43^{+0.03}_{-0.04}$ keV
	measured  in 2006 during quiescence \citep{gelfand07}.
Likewise, the  blackbody radius of $5.2^{+0.4}_{-0.3}(d/9 \textrm{kpc})^2$ km
	from our Model A (table \ref{tab:spec_phase_average}),
	which implies a sizable fraction of a neutron star surface,
	is larger than the values of 1.7--3.7 km measured
	in 2006 August to 2007 August \citep{gelfand07,halpern08}.
Therefore, the enhanced activity caused increases, 
	both in the area and  temperature,
	of the blackbody source located presumably on the neutron star surface.
A qualitatively similar result was obtained by \citet{halpern08},
	utilizing a factor 14 change in the 1--8 keV flux
	from 2006 August to 2007 June.

Since the present Suzaku observation has allowed the first detection 
	of 1E~1547.0$-$5408 in energies above $\sim 10$ keV,
	it is not trivial to evaluate
	how the hard tail was enhanced by the bursting activity.
Let us then use spectra below 10 keV,
	and define a fractional luminosity  carried by any ``non-blackbody" component,
	as $\eta=(L_{\mathrm{tot}} - L_{\mathrm{BB}})/L_{\mathrm{tot}}$,
	where $L_{\mathrm{tot}}$ is the absorbed  1--10 keV total luminosity at 9 kpc,
	and $L_{\mathrm{BB}}$ is the bolometric luminosity of the soft blackbody.
In the present case, we have
	$L_{\mathrm{BB}}=(6.2\pm1.2) \times 10^{35}$ ergs s$^{-1}$ 
	(section~\ref{broadband}),
	together with  $L_{\mathrm{tot}}=1.1 \times 10^{36}$ ergs s$^{-1}$,
	and hence $\eta \sim 0.44$.
Thus, even in energies below 10 keV,
	nearly half the persistent X-ray flux in the present activity 
	is considered to be carried by the hard component,
	unless it steeply cuts off toward lower energies.

The results by  \citet{halpern08},
	obtained in the less active periods in 2006--2007, 
	in contrast imply $\eta=0.26-0.32$.
We therefore infer that 
	the hard tail component is more strongly affected by the activity variation
	than the soft emission.
This inference is reinforced if we notice that
	the ``non-blackbody" fraction in 2006--2007 could be 
	at least partially attributed, e.g., 
	to the Comptonization tail of the soft component itself,
	rather than to  the genuine hard tail component.
In agreement with this inference, \citet{enoto10} suggest
	that the hard component of SGR 0501+4156 decayed,
	after its 2008 August emergence, 
	more rapidly than the soft component.

Yet another effect of the enhanced activity is 
	a marked increase in the pulsed fraction.
In the  less active state in 2007,
	the pulsed fraction was measured with XMM-Newton 
	to be $\sim 7\%$ \citep{halpern08}.
This value was defined, differently from our definition (section~\ref{pulsations}), 
	as a fraction of counts
	above a minimum of the folded pulse profile.
Even using this definition,	
	the 1--3 and 3--10 keV pulsed fraction
	of the present XIS0 data is obtained as 
	$\sim$17\% and $\sim$20\%,
	respectively,
	which are not much different from these given in figure \ref{fig:fractions}.
Thus, 
	the pulsed fraction in the present active sate 	
	is 2--3 times higher than that in the less active one.
Such a change of the pulse fraction between the outburst 
	and quiescence has been reported also from 
	XTE J1810-197 and CXOU J164710.2-455216 \citep{israel07}.
A possible scenario is that these objects are observed
	relatively parallel to their rotational axes,
	and a blackbody emitter localized on the neutron-star surface
	(may or may not be at a magnetic pole) is always visible from us normally.
As the activity gets higher, the blackbody emitter,
	gradually increasing in area  as we found above,
	may start to exhibit self-eclipse as the star rotates.

\section{Summary}


On 2009 January 28--29,
	we performed a SuzakuToO observation of the fastest-rotating magnetar
	1E~1547.0$-$5408 in its burst-active state.

\begin{enumerate}

\item 
In addition to short bursts,
	the persistent X-ray emission was detected over a broad energy band,
	from $\sim 0.8$ keV at least up to $\sim$100 keV.
The 20--100 keV flux of $1.3\times 10^{-10}$ ergs s$^{-1}$ cm$^{-2}$,
	measured for the first time from this source,
	exceeds by a factor of 1.6  (or 2.1) the 
	absorption-corrected  (or uncorrected) 2--10 keV flux.

\item 
The pulsation was detected up to $\sim 70$ keV at a period of 2.072135 s.
Although the pulsed fraction  (16--28\%) 
	is lower than those of most of other magnetars,
	it is considerably higher than was observed
	previously from this object  in quiescence.

\item 
The broad band X-ray spectra were fitted successfully
	by a soft blackbody of $kT=0.65$ keV plus
	an extremely hard power-law component of $\Gamma \sim 1.5$.
At a 9 kpc distance,
	the bolometric luminosity of the former becomes 
$(6.2\pm1.2) \times 10^{35}$  ergs s$^{-1}$, 
	while the 2--100 keV luminosity of the latter
	$1.9\times 10^{36}$ ergs s$^{-1}$.

\item 
Compared to the quiescence,
	the blackbody temperature increased by a factor of $\sim 1.5$,
	and the blackbody radius ($\sim 5$ km at 9 kpc) also increased.
As a result, the 2--10 keV flux incureased by 1--2 order of magnitude
	compared to the previous fainter states in 2006 and 2007.

\item 
Replacing the blackbody with a Comptonized blackbody 
	or an RCS model 
	improved the fit,
	suggesting that the soft component deviates from a pure blackbody.

\item 
The hard X-ray flux, though not measured previously,
	is inferred to have increased during the activity 
	by a larger factor than   the soft X-ray flux.
In the present data, the hard component still carries
	nearly half the flux in energies below 10 keV.

\item 
These results can be understood by considering
	that 1E~1547.0$-$5408, during the  activity,
	became a magnetar that has properties in between those
	of the most active SGRs and of the least active AXPs.

\end{enumerate}

\bigskip
We thank the Suzaku operation team for the successful
	accomplishment of the present ToO observation,
	and are grateful to Tadayasu Dotani for
	his advices on the XIS timing modes.
We also thank Nanda Rea and Gianluca Israel for helpful discussions
	and for providing us with useful information.
In \S4, 
	we used quick-look results provided through a website by the ASM/RXTE team.
TE is supported by the JSPS Research Fellowship for Young Scientists.
The present work was supported in part by Grant-in-Aid for Scientific Research 
	(S), No. 18104004.

\appendix
\section*{Comptonized blackbody Model}
\label{Comptonized blackbody Model}
	When an X-ray photon with an energy $\epsilon'$ is up-scattered 
	in a single Compton process to an energy $\epsilon$,
	its energy becomes $\epsilon_k \sim \epsilon' A^k $ after $k$ scatterings,
	where $ A\equiv  \epsilon/\epsilon' $ denotes an amplification factor.
When the scatterer has an optical depth $\tau$ and is optically thin,
	a probability  for a photon to experience  $k$ scatterings is given as $\tau ^k$.
These scatterings changes the original photon number flux $I(E')$ to
\[
I(E)\sim I(E')\tau^k
=I(E')\left( \frac{E}{E'} \right)^{-\alpha},
\]
where  $\alpha \equiv - \ln \tau/\ln A$.
If the original radiation is a blackbody radiation (in photon number flux),
\[
I(E')=\frac{E'^2}{\exp(E'/kT)-1} \textrm{\ photons s$^{-1}$ cm$^{-2}$ keV$^{-1}$}
\]
with a temperature $kT$ and a normalization factor $C$, 
then the modified spectra after multiple scatterings becomes,
\begin{eqnarray*}
I(E)dE &=& \int^{E}_0 I(E') \left( \frac{E}{E'} \right)^{-\alpha} dE' \quad \mathrm{photons \  s^{-1} \ cm^{-2}} \nonumber \\
I(E) &=& CE^{-1-\alpha} \int^{E}_0 \frac{E'^{2+\alpha}}{\exp(E'/kT)-1}dE'  \\
&=& CE^{-\Gamma_{\mathrm{comp}}} \int^{E}_0 \frac{E'^{\Gamma_{\mathrm{comp}}+1}}{\exp(E'/kT)-1}dE' .
\end{eqnarray*}
Here a photon index in the high energy range is defined as $\Gamma_{\mathrm{comp}}=1+\alpha$,
because it has a form of 
$I(E) \propto E^{-(1+\alpha)} = E^{-\Gamma_{\mathrm{comp}}}$ at $E/kT\gg1$.
This also has the Reyleigh-Jean form in low energy region ($E/kT\ll1$),
$I(E) \propto E$ photons s$^{-1}$ cm$^{-2}$ keV$^{-1}$.
We implemented this integration in {\tt xspec12} 
with the GNU Scientific Library (GSL) codes,
where the integration has a relative accuracy of $10^{-4}$.

\bigskip
\bigskip
\begin{table}[htb]
\caption{Count rates from individual detectors.}
\label{tbl:counts}
\begin{tabular}{lcccc}
\hline \hline
Detector                        &    Energy    & Background- 	      & Background-            & Burst-\\
                                      &     (keV)        & inclusive		      & subtracted                & removed\\
\hline 
XIS1$^*$                       &   2--10        & 1.44$\pm$0.01     & 1.40$\pm$0.01        & 1.34$\pm$0.01\\
XIS3$^*$                       &   2--10        & 1.64$\pm$0.01     & 1.61$\pm$0.01        & 1.53$\pm$0.01\\
\hline 
HXD-PIN$^*$    &  15--70      & 0.528$\pm$0.004 & 0.174$\pm$0.004   & 0.168$\pm$0.004\\
\hline 
HXD-GSO$\dagger$  & 50--114    & 9.40$\pm$0.06      & 0.151$\pm$0.064   & 0.148$\pm$0.064 \\
HXD-GSO$^\dagger$ &114--578  &  27.01$\pm$0.19  & 0.030$\pm$0.192\  & 0.028$\pm$0.192\\
\hline \hline
\end{tabular}\\
$^*$: Counts  s$^{-1}$ with statistical 1$\sigma$ errors. \\
$^\dagger$: Counts s$^{-1}$, with statistical (1$\sigma$) and systematic errors. \\
\end{table}



\begin{table}
\caption{Contaminating source candidates\footnotemark[$*$]}\label{tab:cont_list}
 \begin{center}
    \begin{tabular}{lccc}
      \hline \hline
      Name & Type & Intensity\footnotemark[$\dagger$] & Flux\footnotemark[$\ddagger$]  \\
      \hline
1H 1538-522 &	 HMXB &   9.8 mCrab & $\sim$$1\times 10^{-5}$ \\
H 1608-522   & LMXB  &   19 mCrab & $\sim$$8\times 10^{-5}$ \\
XTE J1550-564 & LMXB & 12 mCrab &  $\sim$$2\times 10^{-4}$ \\
XTE J1543-568 & HMXB  & 0.23 mCrab & $\sim$$4\times 10^{-4}$ \\
Cir X-1 & LMXB & $<$0.1 mCrab & $<$$5\times 10^{-7}$ \\
      \hline
      \multicolumn{4}{@{}l@{}}{\hbox to 0pt{\parbox{85mm}{\footnotesize
\footnotemark[$*$]:  Source  within the FOV of HXD-GSO with $>$1 mCrab in the 20--60 keV range,
from the INTEGRAL General Reference Catalog.       
\par\noindent
	\footnotemark[$\dagger$]: 
Average 2--10 keV intensity, measured by the ASM/RXTE,
over 10 days around the Suzaku observation.
       \par\noindent
       \footnotemark[$\ddagger$]: 
       Predicted flux (photons s$^{-1}$ cm$^{-2}$) of HXD-GSO, 
 using spectral parameters in the INTEGRAL General Reference Catalog, 
 and normalized by  the ASM intensities.
The angular transmission of the HXD-GSO FOV is not considered.
     }\hss}}
    \end{tabular}
  \end{center}
\end{table}

\vspace*{-3cm}

\begin{table}[htb]
\begin{center}
\caption{Phase averaged spectral parameters for each model.$*$}
\label{tab:spec_phase_average}
\begin{tabular}{lccc}
\hline                          
\multicolumn{1}{c}{Model} & (A)  & (B) & (C)  \\
                            &  BB+PL &  CBB+PL & RCS+PL \\
\hline
$N_{\mathrm{H}}$ ($10^{22}$ cm$^{-2}$) &
$3.2\pm0.1$ & $3.3\pm0.1$ & $4.0^{+0.4}_{-0.6}$\\
$kT$ (keV) & 
$0.65\pm0.02$ & $0.48^{+0.05}_{-0.04}$ & --\\
$R$ (km)  \footnotemark[$\dagger$] &
 $5.2^{+0.4}_{-0.3}$ & -- & -- \\
$\alpha=\Gamma_{\mathrm{comp}}-1$ &
-- & $3.9^{+1.9}_{-0.8}$ & -- \\
RCS T (keV) & 
-- & -- & $0.12_{-0.01}^{+0.17}$ \\
RCS $\tau$ & 
-- & -- &  $>$7 \\
RCS $\beta$ & 
-- & -- & $0.49_{-0.19}^{+0.01}$ \\
Absorbed $F_{\mathrm{soft}}$\footnotemark[$\ddagger$] & 
$5.71^{+0.15}_{-0.18}$ & $5.68^{+0.45}_{-0.11}$ & $5.67^{\mathrm{ND}}_{\mathrm{ND}}$ \\
Unabsorb. $F_{\mathrm{soft}}$\footnotemark[$\ddagger$] & 
$7.95^{+0.21}_{-0.25}$ & $7.99^{+0.63}_{-0.15}$ & $7.98^{\mathrm{ND}}_{\mathrm{ND}}$ \\
\hline
$\Gamma_{\mathrm{hard}}$ &
$1.54^{+0.06}_{-0.05}$ & $1.34^{+0.14}_{-0.15}$ & $1.33_{-0.08}^{+0.08}$ \\
$F_{\mathrm{hard}}$ \footnotemark[$\#$] & 
$12.5_{-1.1}^{+0.8}$ & $14.4^{+1.9}_{-1.6}$ & $14.7^{+1.3}_{-1.1}$ \\
\hline
$\chi^2$/d.o.f &  $298.9/278$ & $287.9/277$ & $288.7/276$ \\
                      &$=1.08$    & $ =1.04$ & $ =1.04$ \\
Null hyp. prob. & 
0.19 & 0.31 & 0.32 \\
\hline
      \multicolumn{4}{@{}l@{}}{\hbox to 0pt{\parbox{85mm}{\footnotesize
\par\noindent
\footnotemark[$*$]: 
BB, PL, CBB, and RCS represent
 blackbody, power-law, (magnetic) comptonized blackbody, and 
resonant cyclotron scattering, respectively.
ND represents ``not determined."
All the quoted errors are only statistical at the 90\% confidence level.
\par\noindent
\footnotemark[$\dagger$]: 
Blackbody radius assuming a distance of 9 kpc.
\par\noindent
\footnotemark[$\ddagger$]:  
2--10 keV  fluxe in $10^{-11}$ ergs\,s$^{-1}$\,cm$^{-2}$ .
\par\noindent
\footnotemark[$\#$]: 
Unabsorbed 20--100 keV flux in $10^{-11}$ ergs\,s$^{-1}$\,cm$^{-2}$.
 } \hss}}
\end{tabular}
\end{center}
\end{table}

\clearpage

\end{document}